\documentclass[prd,twocolumn,showpacs,amsmath,amssymb,superscriptaddress,floatfix,nofootinbib]{revtex4}
\usepackage{mathrsfs,bm}
\usepackage{longtable,lscape}
\usepackage{txfonts}
\usepackage{amssymb}
\usepackage{indentfirst}
\usepackage{graphicx,,booktabs}
\usepackage{ulem}
\usepackage{multirow}
\usepackage{color}
\usepackage{amssymb}
\definecolor{cover}{rgb}{0.77,0.87,0.88}
\definecolor{blueone}{rgb}{0.1,0.1,.7}
\definecolor{citec}{rgb}{0.14,0.47,0.09}
\definecolor{two}{rgb}{0.0,0.5,0.}
\definecolor{three}{rgb}{.5,.1,0.15}
\usepackage[bookmarks=true,bookmarksopen=false,plainpages=false,breaklinks=true,
   bookmarksnumbered=true,hypertexnames=false,filecolor=blue,urlcolor=blueone,menucolor=three,
   linkcolor=three,citecolor=blueone, colorlinks,anchorcolor=blue,%
   pdfstartview=FitH,pdftitle=title,%
   pdfauthor=author]{hyperref}

\usepackage{relsize}
\usepackage{xspace}
\def\babar{\mbox{\slshape B\kern-0.1em{\smaller A}\kern-0.1em
    B\kern-0.1em{\smaller A\kern-0.2em R}}}

\begin{document}
\title{Production of the $D_{s0}(2317)$ and $D_{s1}(2460)$ by kaon-induced reactions on a proton target}

\author{HongQiang Zhu}
\affiliation{College of Physics and Electronic Engineering, Chongqing Normal University,  Chongqing 401331,China}

\author{Yin Huang \footnote{corresponding author}}
\email{huangy2017@buaa.edu.cn}
\affiliation{School of Physics and Nuclear Energy Engineering, Beihang University, Beijing 100191, China}

\date{\today}
\begin{abstract}
We investigate the possibility to study the charmed-strange mesons $D_{s0}(2317)$ and $D_{s1}(2460)$ by
kaon-induced reactions on a proton target in an effective Lagrangian approach.   The production
process is described by the $t$-channel $D^{0}$ and $D^{*0}$ exchanges, respectively.  Our theoretical
approach is based on the chiral unitary theory where the $D_{s0}(2317)$ and $D_{s1}(2460)$
resonances are dynamically generated.  Within the coupling constants of the $D_{s0}(2317)$ to $KD$
and $D_{s1}(2460)$ to $KD^{*}$ channels obtained from the chiral unitary theory, the total and differential
cross sections of the $K^{-}p\to{}\Lambda_cD_{s0}(2317)$ and $K^{-}p\to{}\Lambda_cD_{s1}(2460)$
are evaluated.    The $\bar{K}p$ initial state interaction mediated by  Pomeron and Reggeon exchanges
is also included, which reduces the production cross sections of the $D_{s0}(2317)$ and $D_{s1}(2460)$. If measured in future experiments,
the predicted total cross sections and specific features of the angular distributions can be used to test
the (molecular) nature of the $D_{s0}(2317)$ and $D_{s1}(2460)$.
\end{abstract}


\maketitle
\section{INTRODUCTION}
In recent years, many charmed-strange mesons have been observed~\cite{Tanabashi:2018oca}.  Among them, the $D_{s0}(2317)$
and $D_{s1}(2460)$ are two peculiar states(we abbreviate it as $D_{s0}^{*}$ and $D_{s1}^{*}$ hereafter), since their masses
are about 160 MeV and 70 MeV, respectively, below the quark model predicted values~\cite{Godfrey:1985xj}.
The charmed-strange meson $D^{*}_{s0}$ was first observed by the BaBar collaboration as a narrow peak in
the $D_{s}\pi$ invariant mass distribution~\cite{Aubert:2003fg}.  The state was shortly confirmed by CLEO~\cite{Besson:2003cp}
and Belle~\cite{Abe:2003jk,Krokovny:2003zq} collaborations.  Nowadays it is  well established in the PDG~\cite{Tanabashi:2018oca}
with quantum numbers $I(J^{P})=0(0^{+})$.  The $D^{*}_{s1}$ was also observed in the CLEO experiment~\cite{Besson:2003cp}
in the $D_s^{*}\pi$ channel and BaBar~\cite{Aubert:2004pw,Aubert:2003pe,Aubert:2006bk} also found a signal in that
region. Nowadays it is also well established  in the PDG~\cite{Tanabashi:2018oca} with quantum numbers $I(J^P)=0(1^{+})$.
The mass and width of the $D_{s0}^{*}$ and $D_{s1}^{*}$ states reported by the above collaborations~\cite{Aubert:2003fg,Besson:2003cp,Abe:2003jk,Krokovny:2003zq,Aubert:2004pw,Aubert:2003pe,Aubert:2006bk}
are consistent with each other, i.e.,
 \renewcommand{\arraystretch}{1.5}
\begin{center}
\begin{tabular}{rcl}
$D_{s0}(2317)^{\pm}$:     \ \  &M&=2317.7$\pm$1.3  ~ {MeV}\nonumber,\\
                            &$\Gamma$&$<$3.8    ~{MeV};\nonumber\\
$D_{s1}(2460)^{\pm}$:  \  \ &M&=2459.5$\pm0.6$  ~ {MeV},\nonumber\\
                            &$\Gamma$&$<$$3.5$  ~ {MeV}.\nonumber\\
\end{tabular}
\end{center}

The large disagreement between the quark model expectations~\cite{Godfrey:1985xj} and the experimental measurements~\cite{Aubert:2003fg,Besson:2003cp,Abe:2003jk,Krokovny:2003zq,Aubert:2004pw,Aubert:2003pe,Aubert:2006bk}
made it difficult to assign these two states as  conventional charmed-strange mesons.   Since the masses of the $D_{s0}^{*}$ and
$D_{s1}^{*}$ are about 40 MeV below the $DK$ and $D^{*}K$ thresholds, respectively, many studies proposed that
the $D_{s0}^{*}$ and $D_{s1}^{*}$  are  $S$-wave $DK$ and $D^{*}K$ molecular states.  The studies in the Bethe-Salpeter
approach~\cite{Xie:2010zza} and potential model~\cite{Zhang:2006ix} showed indeed that the $D^{*}_{s0}$ could be a $DK$ hadronic
molecule.  In Ref.~\cite{Bicudo:2004dx}, the $D_{s0}^{*}$ and $D^{*}_{s1}$ were considered as kaonic molecules bound by strong
short range attraction.  Assuming that the $D_{s0}^{*}$ and $D_{s1}^{*}$ are $DK$ and $D^{*}K$ molecular states,
the strong and radiative decays of the $D_{s0}^{*}$ and $D_{s1}^{*}$ were studied  by several groups~\cite{Cleven:2014oka,Faessler:2007us,Faessler:2007gv,Xiao:2016hoa}.
The production of the $D_{s0}^{*}$ and $D_{s1}^{*}$
from the nonleptonic $B$ decay were calculated in Ref.~\cite{Datta:2003re},  in which the $D_{s0}^{*}$ and $D_{s1}^{*}$ were also
considered as hadronic molecules of $DK$ and $D^{*}K$, respectively.  In the chiral unitary
approach~\cite{Gamermann:2006nm,Gamermann:2007fi,Altenbuchinger:2013vwa,MartinezTorres:2018zbl,Guo:2006fu,Guo:2006rp},
the $D_{s0}^{*}$ and $D_{s1}^{*}$ can be dynamically generated from the $DK/D^*K$ and coupled channel interactions.

In addition to the interpretation of the $D_{s0}^{*}$ and $D_{s1}^{*}$ as $DK$ and $D^{*}K$ molecules,
the possibility to assign them as a conventional open charmed meson was also discussed in many different  approaches, such as the relativistic
quark model~\cite{jb:2014ms}, the chiral perturbation theory~\cite{Fajfer:2015zma}, the quark pair-creation model~\cite{Liu:2006jx,Lu:2006ry},
and the QCD sum rules~\cite{Wang:2006mf,Dai:2003yg,Colangelo:2003vg,Colangelo:2005hv,Colangelo:2012xi}.  On the other hand, the large-$N_c$
expansion indicated that the $D_{s0}^{*}$ could be a tetraquark meson~\cite{Guo:2015dha}.   The tetraquark interpretation
was also proposed to understand the mass and decay behavior of the $D_{s0}^{*}$~\cite{Nielsen:2005zr}.  We note that
the QCD sum rules also supported the idea that the $D_{s0}^{*}$ does not seem to be a standard quark-antiquark meson~\cite{Wang:2006uba}.

The present knowledge about the $D_{s0}^{*}$ and $D_{s1}^{*}$ was obtained from the $e^{+}e^{-}$
collision~\cite{Aubert:2003fg,Besson:2003cp,Abe:2003jk,Krokovny:2003zq,Aubert:2004pw,Aubert:2003pe,Aubert:2006bk}.  Thus,
it will be helpful to understand the nature of the $D_{s0}^{*}$ and $D_{s1}^{*}$ if we can observe them in other production
processes.  High-energy kaon beams are available at OKA@U-70~\cite{Obraztsov:2016lhp} and SPS@CERN~\cite{Velghe:2016jjw},
which provide another alternative to study $D_{s0}^{*}$ and $D_{s1}^{*}$.  The kaon beam at J-PARC can also be  upgraded
to the energy region required in charmed-strange meson productions~\cite{Nagae:2008zz}.  Therefore, it is interesting to study the
$D_{s0}^{*}$ and $D_{s1}^{*}$ productions in the $K^{-}p\to{}\Lambda_cD_{s0}^{*}$ and $K^{-}p\to{}\Lambda_cD_{s1}^{*}$
reactions.   Since there exist plenty of experimental information about the $Kp$ elastic interaction in the energy region
relevant to the $D_{s0}^{*}$ and $D_{s1}^{*}$ production~\cite{Aubert:2003fg,Besson:2003cp,Abe:2003jk,Krokovny:2003zq,Aubert:2004pw,Aubert:2003pe,Aubert:2006bk},
the effect from the $Kp$ initial state interaction (ISI) can be taken into account in order to make a more reliable prediction.

This paper is organized as follows.  In Sec. II, we will present the theoretical formalism.  In Sec. III,  the
numerical result of the kaon-induced $D_{s0}^{*}$ and $D_{s1}^{*}$ production on a proton target will be given,
followed by discussions and conclusions in the last section.

\section{Theoretical FORMALISM}
\begin{figure}[h!]
\begin{center}
\includegraphics[bb=90 600 650 710, clip,scale=0.60]{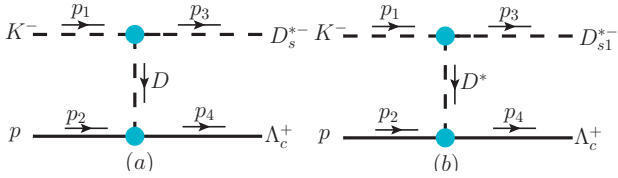}
\end{center}
\caption{Feynman diagram for the mechanism of the $D_{s0}^{*}$ and $D_{s1}^{*}$
production in the $K^{-}p\to{}D^{*-}_{s0}\Lambda_c$ and $K^{-}p\to{}D^{*-}_{s1}\Lambda_c$
reaction.  We also show the definition of the kinematics ($p_1$,~
$p_2$,~$p_3$,and~$p4$) used in the calculation.  }\label{feydiagrams}
\end{figure}

The tree level Feynman diagrams for the $K^{-}p\to{}\Lambda_cD^{*}_{s0}$ and $K^{-}p\to{}\Lambda_cD^{*}_{s1}$
reactions are depicted in Fig.~\ref{feydiagrams}, where the $t-$channel $D$ and $D^{*}$ exchange are considered.
 In this work, the contributions from $s-$ and $u-$ channels are ignored, because the $s-$ and $u-$channels,  which
 involves the creation of an additional $s\bar{s}$ quark pair creation  in the kaon-induced production, are usually
 strongly suppressed,   Hence, the $K^{-}p\to{}\Lambda_cD^{*}_{s0}$ and $K^{-}p\to{}\Lambda_cD^{*}_{s1}$ reactions should be dominated by
the Born terms through the $t$-channel $D$ and $D^{*}$ exchanges, which makes the background very small.

\subsection{Lagrangians}
To compute the diagrams shown in Fig.~\ref{feydiagrams}, we need the effective Lagrangian densities
for the relevant interaction vertices.  For the $\Lambda_cpD$ and $\Lambda_cpD^{*}$ vertices, we adopt the commonly
employed Lagrangian densities as follows~\cite{Dong:2014ksa},
\begin{align}
&{\cal{L}}_{\Lambda_cpD}=ig_{\Lambda_cpD}\bar{\Lambda}_c\gamma_5pD^{0}+{\rm H.c.},\\
&{\cal{L}}_{\Lambda_cpD^{*}}=g_{\Lambda_cpD^{*}}\bar{\Lambda}_c\gamma^{\mu}pD^{*0}_{\mu}+{\rm H.c.}.
\end{align}
The coupling constants $g_{\Lambda_cpD}=-13.98$ and $g_{\Lambda_cpD^{*}}$=-5.20 are determined from the SU(4)
invariant Lagrangians~\cite{Dong:2010xv} in terms of $g_{\pi{}NN}$=13.45 and $g_{\rho{}NN}$=6.0.

In addition to the $\Lambda_cpD$ and $\Lambda_cpD^{*}$ vertices, we also need the information on the $KDD^{*}_{s}$
and $KD^{*}D^{*}_{s1}$ vertices.  As mentioned in the chiral unitary approach of Refs.~\cite{Gamermann:2006nm,Gamermann:2007fi,Altenbuchinger:2013vwa,MartinezTorres:2018zbl,Guo:2006fu,Guo:2006rp},
the $D_s^{*}$ and $D_{s1}^{*}$ resonance are identified as $s-$wave meson-meson molecule that include big $\bar{K}D$ and $\bar{K}D^{*}$
component, respectively.   We can write down the $KDD^{*}_{s}$ and $KD^{*}D^{*}_{s1}$ vertices of Fig.~\ref{feydiagrams} as
\begin{align}
&{\cal{L}}_{KDD^{*}_{s0}}=g_{KDD^{*}_{s0}}\bar{K}DD^{*}_{s0},\\
&{\cal{L}}_{KD^{*}D^{*}_{s1}}=g_{KD^{*}D^{*}_{s1}}\bar{K}D^{*\mu}D^{*}_{s1,\mu},
\end{align}
where the coupling of the $D^{*}_{s0}$ to $D^{0}K^{-}$, $g_{KDD^{*}_{s0}}$, is obtained from the coupling constant of the
$D^{*}_{s0}$ to the $DK$ channel in isospin $I$ = 0, found to be $g_{KDD^{*}_{s}}=10.21$ in Ref.~\cite{Gamermann:2006nm},
multiplied by the appropriate Clebsch-Gordan (CG) coefficient, namely $g_{K^{-}D^{0}D^{*-}_{s}}=g_{KDD^{*}_{s}}/\sqrt{2}$.
Similarly to  Ref.~\cite{Gamermann:2006nm}, we rely on the chiral unitary approach~\cite{Gamermann:2007fi}
to obtain the coupling constant $g_{K^{-}D^{*0}D^{*-}_{s1}}=9.82/\sqrt{2}$.

When evaluating the scattering amplitudes of the $K^{-}p\to{}\Lambda_cD^{*}_s$ and $K^{-}p\to{}\Lambda_cD^{*}_{s1}$ reactions,
we need to include form factors because  hadrons are not point-like particles.  We adopt here a common scheme used in
many previous works~\cite{He:2011jp,Xie:2015zga},
\begin{align}
F_{D^{(*)}}(q^2_{D^{(*)}},M_{ex})=\frac{\Lambda_{D^{*}}^2-M_{D^*}^2}{\Lambda_{D^{*}}^2-q_{D^*}^2},
\end{align}
for the $t$-channel $D^{(*)}$ meson exchange.   Here the $q_{D^{(*)}}$ and $M_{D^{(*)}}$ are the four momentum and the mass of the
exchanged $D^{(*)}$ meson, respectively.  In this model, the $\Lambda_{D^{(*)}}$ is the hard cutoff and it  can be directly
related to the hadron size.  Empirically, the cutoff parameter $\Lambda_{D^{(*)}}$ should be at least a few hundred MeV larger than the $D^{(*)}$ mass.
Hence, we chose $\Lambda_{D^{(*)}}=M_{D^{(*)}}+\alpha\Lambda_{QCD}$ with $\Lambda_{QCD}=0.22$ GeV as used in previous
works~\cite{Huang:2013mua,Xiao:2017uve,Xu:2015qqa} for other reactions.   The parameter $\alpha$ is usually close to unitary, and in this work a variation
of the cutoff will be made to show the sensitivity of the results on the cutoff.

\subsection{Initial state interaction(ISI)}
\begin{figure}[h!]
\begin{center}
\includegraphics[bb=90 610 600 710, clip,scale=0.60]{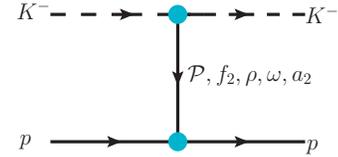}
\end{center}
\caption{Feynman diagram for the mechanism of the initial state interaction of the $K^{-}p$ system.}\label{isi}
\end{figure}

Following Ref.~\cite{Lebiedowicz:2011tp}, the initial state interaction for the $K^{-}p\to{}K^{-}p$ reaction at high energies will
be taken into account and the relevant Feynman diagram for the Initial state interaction(ISI) is shown in Fig.~\ref{isi}.
The amplitude ${\cal{T}}_{K^{-}p\to{}K^{-}p}$ is written in terms of the Pomeron and Reggeon exchanges~\cite{Lebiedowicz:2011tp}
\begin{align}
{\cal{T}}_{K^{-}p\to{}K^{-}p}=A_{{I\!\!P}}+A_{f_2}+A_{a_2}+A_{\omega}+A_{\rho}\label{qwq}.
\end{align}
When the center-of-mass energy $\sqrt{s}$ is large, the elastic $K^{-}p$ scattering amplitude is a sum of the following terms,
\begin{align}
A_{i}(s,t)=\eta_isC_{i}^{KN}(\frac{s}{s_0})^{\alpha_i(t)-1}\exp(\frac{B^{i}_{KN}}{2}t),
\end{align}
where $i= I\!\!P$ for Pomeron and $f_2$, $a_2$, $\omega$, and $\rho$  Reggeons.  The energy scale $s_0=1$ GeV$^2$.  The  coupling
constants $C_i^{KN}$, the parameters of the Regge linear trajectories[$\alpha_i(t)=\alpha_i(0)+\alpha^{'}_it$], the signature
factors($\eta_i$), and the $B^{i}_{KN}$ used in Ref.~\cite{Lebiedowicz:2011tp} provide a rather good description of the experimental
data.  The parameters determined in Ref.~\cite{Lebiedowicz:2011tp}  are listed in Table.~\ref{table-coff}.
\begin{table}[h!]
\centering
\caption{Parameters of Pomeron and Reggeon exchanges determined from elastic and total cross sections in
Ref.~\cite{Lebiedowicz:2011tp}.}\label{table-coff}
\begin{tabular}{ccccc}
\hline\hline
$i$       ~~&    $\eta_i$    & ~~$\alpha_i(t)$               & ~~$C_i^{KN}$(mb)      &~~$B_i^{KN}$ (GeV$^{-2})$   \\ \hline
$I\!\!P$     ~~&      $i$       & ~~1.081+(0.25 GeV$^{-2}$)t    & ~~11.82               &~~2.5                        \\
$f_2$     ~~& $-0.861+i$     & ~~0.548+(0.93 GeV$^{-2}$)t    & ~~15.67               &~~2.0                        \\
$\rho$    ~~& $-1.162-i$     & ~~0.548+(0.93 GeV$^{-2}$)t    & ~~2.05                &~~2.0                        \\
$\omega$  ~~& $-1.162-i$     & ~~0.548+(0.93 GeV$^{-2}$)t    & ~~7.055               &~~2.0                        \\
$a_2$     ~~& $-0.861+i$     & ~~0.548+(0.93 GeV$^{-2}$)t    & ~~1.585               &~~2.0                        \\   \hline \hline
\end{tabular}
\end{table}

\section{Kaon-induced $D_{s}^{*}$(2317) and $D^{*}_{s1}$(2460) production on proton target}
First, we calculate  the total cross section of the $K^{-}p\to{}\Lambda_c^{+}D^{*-}_s$ and $K^{-}p\to{}\Lambda_c^{+}D^{*-}_{s1}$
reactions   The corresponding unpolarized differential cross section reads
\begin{align}
\frac{d\sigma}{d\cos\theta}=\frac{m_pm_{\Lambda_c}}{8\pi{}s}\frac{|\vec{p}_{3cm}|}{|\vec{p}_{1cm}|}(\frac{1}{2}\sum_{s_c,s_2}|{\cal{M}}^{1/2^{\pm}}|^2),
\end{align}
where  $s=(p_1+p_2)^2$,  $\theta$ is the scattering angle of the outgoing meson relative to the beam direction,
$\vec{p}_{1cm}$ and $\vec{p}_{3cm}$ are the $K^{-}$ and $D_{s}^{*-}$($D_{s1}^{*-}$) three momenta in the center
of mass frame,
\begin{align}
&|\vec{p}_1|=\frac{\lambda^{1/2}(s,m^2_{K^{-}},m_p^2)}{2\sqrt{s}},\\
&|\vec{p}_3|=\frac{\lambda^{1/2}(s,m^2_{D_{s/s1}^{-*}},m_{\Lambda^{+}_c}^2)}{2\sqrt{s}},
\end{align}
where $\lambda(x,y,z)$ is K$\ddot{a}$llen function with $\lambda(x,y,z)=(x-y-z)^2-4yz$. The $m_{K^{-}}$, $m_p$, and $m_{\Lambda_c}$
are the masses of the $K^{-}$ meson, proton, and $\Lambda_c$, respectively.  Here, we take $m_{K^{-}}=493.68$ MeV, $m_p=938.27$ MeV,
and $m_{\Lambda_c}=2286.46$ MeV.

Taking the ISI of the $K^{-}p$ system into account, the full amplitude for the process $K^{-}p\to\Lambda_c^{+}D^{*-}_{s0/s1}$ is a sum
of the Born and ISI amplitudes.   With the Lagrangians given in the previous section,  the Born amplitude of the
$K^{-}(p_1)p(p_2)\to{}\Lambda_c^{+}(p_4)D^{*-}_{s0/s1}(p_3)$ reaction can be obtained as,
\begin{align}
{\cal{M}}_B^{D^{*}_{s0}}&=-g_{\Lambda_cpD^{*}_{s0}}\bar{u}(p_4,s_c)\gamma_{5}u(p_2,s_2)\frac{1}{(p_3-p_1)^2-m^2_{D}}\nonumber\\
                         &\times{}g_{KD^{*}_{s0}D}F^2_{D}((p_3-p_1)^2,m_{D}),\\
{\cal{M}}_B^{D^{*}_{s1}}&=ig_{\Lambda_cpD^{*}_{s1}}\bar{u}(p_4,s_c)\gamma_{\mu}u(p_2,s_2)\frac{1}{(p_3-p_1)^2-m^2_{D^{*}}}\nonumber\\
                         &\times{}(-g^{\mu\nu}+\frac{(p_3-p_1)^{\mu}(p_3-p_1)^{\nu}}{m^2_{D^{*}}})\nonumber\\
                         &\times{}g_{KD^{*}_{s1}D^{*}}F^2_{D}((p_3-p_1)^2,m_{D^{*}})\epsilon^{*}_{\nu}(p_3),
\end{align}
where $\bar{u}(p_4,s_c)$ and $u(p_2,s_2)$ are the Dirac spinors with $s_c$ ($p_4$) and $s_2$ ($p_2$) being the spins (the four-momenta)
of the outgoing $\Lambda_c$ and the initial proton, respectively.  The $\epsilon^{*}_{\nu}(p_3)$ is the polarization vector of the $D^{*}_{s1}$.

Following the strategy of Ref.~\cite{Lebiedowicz:2011tp}, the ISI amplitude can be written as
\begin{align}
{\cal{M}}^{D^{*}_{s/s1}}_{ISI}=\frac{i}{16\pi^2s}\int{}d^2\vec{k}_t&{\cal{T}}_{K^{-}p\to{}K^{-}p}(s,k^2_t)\nonumber\\
                           &\times{\cal{M}}^{1/2^{\pm}}_{D^{*}_{s/s1}}(-p_2-k_t+p_4),
\end{align}
where  $k_t$ is the momentum transfer in the $K^{-}p\to{}K^{-}p$ reaction.

\begin{figure}[h!]
\begin{center}
\includegraphics[bb=60 10 800 420, clip,scale=0.33]{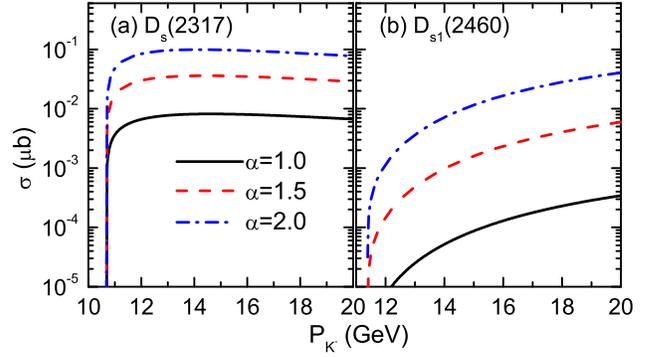}\\
\caption{(color online). Total cross section $\sigma$ for the
$K^{-}p\to{}\Lambda_c^{+}D^{*-}_{s0}$(a) and $K^{-}p\to{}\Lambda_c^{+}D^{*-}_{s1}(b)$ reaction as a
function of the beam momentum $p_{K^{-}}$.} \label{cross22}
\end{center}
\end{figure}

With the formalism and ingredients given above, the total cross section versus the beam momentum of the $K^{-}p$ system for the
$K^{-}p\to{}\Lambda_c^{+}D^{*-}_{s0}$ and $K^{-}p\to{}\Lambda_c^{+}D^{*-}_{s1}$ reactions are evaluated.  The results for beam momentum
$P_{k^{-}}$ from the reaction threshold to 20.0 GeV are shown in Fig.~\ref{cross22}.   Because the cutoff can not be well determined,
the results obtained with several cutoffs are also presented.  It is worthy mentioning that the value of the cross section is very
sensitive to the model parameter $\alpha$ when varying the cutoff parameter $\alpha$ from 1.0 to 2.0.  This is because the model parameters
we selected are very close to the masses of the exchanged particles.  To make a reliable prediction for the cross section of the
$K^{-}p\to{}\Lambda_c^{+}D^{*-}_{s0}$ and $K^{-}p\to{}\Lambda_c^{+}D^{*-}_{s1}$ reactions thus requires a good knowledge on the form factors.
More and accurate experimental data can also be used to constrain the value of the cutoff parameter.

The results in Fig.~\ref{cross22} also show that the total cross section increases sharply near the $D^{*-}_{s0/s1}\Lambda_c$ threshold.
At higher energies, the cross section increases continuously but relatively slowly compared with that near threshold.  However, the total
cross section decreases but very slowly for the $D_{s0}^{*}$ production in the $K^{-}p\to{}\Lambda_c^{+}D^{*-}_{s1}$ reaction when we
change the beam energy $P_{K^{-}}$ from 14.6 GeV to 20.0 GeV.    With the increase of the cutoff, the total cross section increases.
Comparing the cross section of the  $K^{-}p\to{}\Lambda_c^{+}D^{*-}_{s0}$ reaction with that of the $K^{-}p\to{}\Lambda_c^{+}D^{*-}_{s1}$
reaction,  we found that the line shapes of the cross section are very different.   A possible explanation for this may be that $KD$
interaction to form the $D^{*}_{s0}$ is stronger than that $KD^{*}$ interaction to form $D^{*}_{s1}$ due to the $D^{*}$ meson decays
completely to the final state containing the $D$ meson~\cite{Tanabashi:2018oca}.

The results show that the total cross section for $D^{*}_{s0}$ production is bigger than that for $D^{*}_{s1}$ production.  At a beam
momentum about 14.6 GeV and the parameter $\alpha=1$ the cross section is of the order of 10 nb, which is quite large for an experimentally
observation of the $D^{*}_{s0}$ at current and future facilities.  Our results suggest that it will takes high energy, at least above 14.6 GeV,
to observe the production of $D^{*}_{s1}$ in the $K^{-}p\to{}\Lambda_c^{+}D^{*-}_{s1}$ reaction.

\begin{figure}[h!]
\begin{center}
\includegraphics[bb=70 5 800 420, clip,scale=0.35]{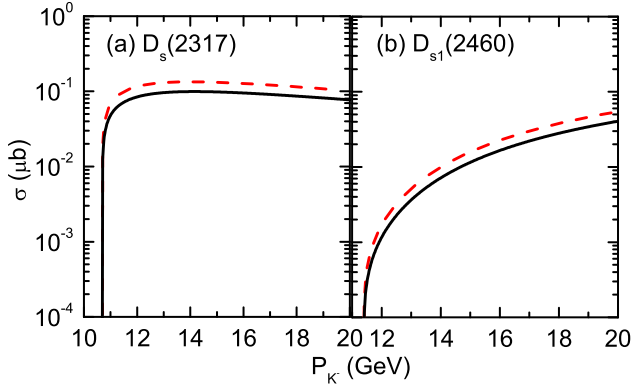}\\
\caption{(color online). Total cross section  with or without ISI for the $K^{-}p\to{}D_{s0/s1}^{*-}
\Lambda_c^{*}$ reaction as a function of the beam momentum $p_{K^{-}}$
for the  $D^{*}_{s0}$ case (left) and the $D^{*}_{s1}$ case (right).} \label{compare23}
\end{center}
\end{figure}

To show the effect from the $K^{-}p$ ISI, we compare the cross sections obtained with and without ISI for the cutoff of $\alpha=2.0$ in
Fig.~\ref{compare23},   for  he $K^{-}p\to{}\Lambda_c^{+}D^{*-}_{s0}$ (panel A) and
$K^{-}p\to{}\Lambda_c^{+}D^{*-}_{s1}$ (panel B) reactions, respectively.   In Fig.~\ref{compare23}, the dashed red lines are the pure Born amplitude
contribution, while the solid black lines are the full results.  It shows that the role of the ISI is to reduce the cross
section by approximately $20\%$, in agreement with the conclusions drawn from Refs.~\cite{Dong:2014ksa,Hanhart:2003pg,Baru:2002rs} that
the ISI for $pp$ or $p\bar{p}$ reactions reduces the cross section.
\begin{figure}[h!]
\begin{center}
\includegraphics[bb=-30 50 500 535, clip,scale=0.43]{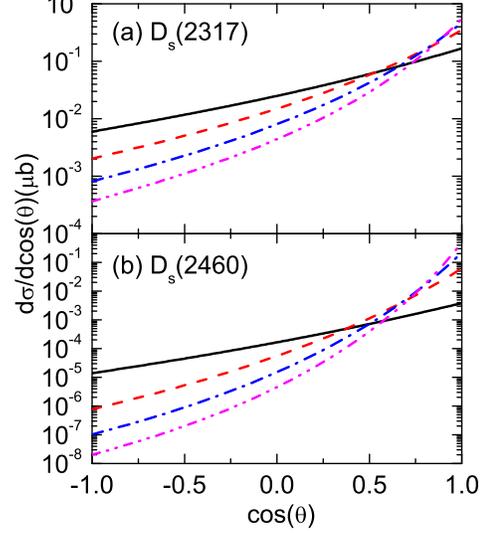}\\
\caption{(color online). The $K^{-}p\to{}\Lambda_c^{+}D^{*-}_{s0}$(a) and $K^{-}p\to{}\Lambda_c^{+}D^{*-}_{s1}$(b)
differential cross section at different energies with $\alpha=2.0$.  The black solid lines, red dash lines,
blue dash dot lines, and straight dash dot dot lines are obtained at beam energy $P_{K^{-}}$=12.0,14.0,16.0, and
18.0 GeV, respectively.} \label{diff-cr22}
\end{center}
\end{figure}

In addition to the total cross section, we also compute the differential cross section for the $K^{-}p\to{}\Lambda_c^{+}D^{*-}_{s0}$
and $K^{-}p\to{}\Lambda_c^{+}D^{*-}_{s1}$ reactions as a function of the scattering angle of the outgoing meson relative to the beam direction at different
beam energies,i.e,$P_{K^{-}}$=12.0,14.0,16.0, and 18.0 GeV.  The theoretical results are shown in Fig.~\ref{diff-cr22}.  We note that the differential cross section is the largest at the extreme forward angle and decreases with the increase of the scattering
angle.  This is because we have only considered the contributions from the $t$-channel $D$ and $D^{*}$ exchanges. It should
be pointed out that, if there are contributions from $s-$ and $u-$ channels,  there will be a clear bump (or peak) in the total cross section
which can be distinguished easily.

\section{SUMMARY}
In this work,  the production of the $D^{*-}_{s0}$ and $D^{*-}_{s1}$ resonances in the $K^{-}p\to{}\Lambda_c^{+}D^{*-}_{s0}$
and $K^{-}p\to{}\Lambda_c^{+}D^{*-}_{s1}$ reactions was studied in  an effective Lagrangian approach.
The production process is described by the $t$-channel $D^{0}$ and $D^{*0}$ meson exchanges, respectively.  The coupling
constants of  the $D_{s0}^{*}$ to $KD$ and $D^{*}_{s1}$ to $KD^{*}$  are obtained from the  chiral unitary
theory~\cite{Gamermann:2006nm,Gamermann:2007fi}. where the $D^{*-}_{s0}$ and $D^{*-}_{s1}$ resonance are dynamically generated.
The $K^{-}p$ initial state interaction(ISI) was included by Pomeron and Reggeon exchanges~\cite{Lebiedowicz:2011tp}, which
was shown to reduce the cross section by about $20\%$.  the total and differential cross sections computed can be used to test the
molecular picture of the $D^{*}_{s0}$ and $D^{*}_{s1}$ mesons in facilities such as OKA@U-70,SPS@CERN,and future J-PARC.

Finally, we would like to stress that, thanks to the absence of the $s-$channel, $u-$channel, and background contribution in
the $K^{-}p\to{}\Lambda_c^{+}D^{*-}_s$ and $K^{-}p\to{}\Lambda_c^{+}D^{*-}_{s1}$ reactions, future experimental
data for these two reactions can be used to improve our knowledge of $D^{*-}_{s0}$ and $D^{*-}_{s1}$ properties, which are at
present poorly known.

\begin{acknowledgments}
Y.H.thank Li-Sheng Geng for valuable discussions.  This work was supported by the Science and Technology Research Program of
Chongqing Municipal Education Commission (Grant No. KJQN201800510), the Opened Fund of the State Key Laboratory on Integrated
Optoelectronics (GrantNo. IOSKL2017KF19), and China Postdoctoral Science Foundation(No.2018M641143).  This work also
was supported in part by the National Natural Science Foundation of China under Grants No. 11522539 and No. 11735003.

\end{acknowledgments}

\end{document}